\documentclass[proof]{pasj00}
\SetRunningHead{Mori et al.}{Near- to Mid- Infrared Imaging+Spectroscopy of the Nearby Merging Galaxy NGC~6240}

\usepackage{times}
\usepackage{ulem}

\begin{document}

\title{Near- to Mid- Infrared Imaging and Spectroscopy of Two Buried AGNs of the Nearby Merging Galaxy NGC~6240 with Subaru/IRCS+AO and GTC/CanariCam}
\author{Tamami~I.~Mori\altaffilmark{1},
            Masatoshi~Imanishi\altaffilmark{2,3,4}, 
            Almudena~Alonso-Herrero\altaffilmark{5},
            Chris~Packham\altaffilmark{6}, 
            Cristina~Ramos Almeida\altaffilmark{7,8}, 
            Robert~Nikutta\altaffilmark{9}, 
            Omaira~Gonz\'alez-Mart\'{\i}n\altaffilmark{7,8}, 
            Eric~Perlman\altaffilmark{10}, 
            Yuriko~Saito\altaffilmark{4}, \&
            N.~A.~Levenson\altaffilmark{11}}
\altaffiltext{1}{Department of Astronomy, Graduate School of Science, The University of Tokyo, 7-3-1 Hongo, Bunkyo-ku, Tokyo 113-0033, Japan}
\altaffiltext{2}{Subaru Telescope, 650 North A'ohoku Place, Hilo Hawaii 96720, U.S.A. }
\altaffiltext{3}{National Astronomical Observatory of Japan, 2-21-1 Osawa, Mitaka, Tokyo 181-8588}
\altaffiltext{4}{Department of Astronomy, School of Science, Graduate University for Advanced Studies (SOKENDAI), Mitaka, Tokyo 181-8588}
\altaffiltext{5}{Instituto de Fisica de Cantabria, CSIC-UC, 39005 Santander, Spain and Augusto G. Linares Senior Research Fellow}
\altaffiltext{6}{Department of Physics and Astronomy, University of Texas at San Antonio, One UTSA Circle, San Antonio, TX 78249, USA}
\altaffiltext{7}{Instituto de Astrof\'{\i}sica de Canarias, Calle V\'{\i}a L\'actea, s/n, E-38205, La Laguna, Tenerife, Spain}
\altaffiltext{8}{Departamento de Astrof\'{\i}sica, Universidad de La Laguna, E-38205, La Laguna, Tenerife, Spain}
\altaffiltext{9}{Universidad Andr\'es Bello, Departamento de Ciencias F\'{\i}sicas, Rep\'ublica 252, Santiago, Chile}
\altaffiltext{10}{Florida Institute of Technology, Melbourne, FL 32901, USA}
\altaffiltext{11}{Gemini Observatory, Casilla 603, La Serena, Chile}

\email{morii@astron.s.u-tokyo.ac.jp}

\maketitle

\begin{abstract}
We report near-infrared {\it K'}, {\it L'}, and {\it M'} band imaging observations of the nearby merging galaxy NGC~6240 
with the Infrared Camera and Spectrograph on the {\it Subaru telescope}. 
The observations were performed with the assistance of the Subaru Adaptive Optics System,
and the achieved spatial resolutions were around 0.10--0.20$^{\prime\prime}$. 
We also obtained new mid-infrared imaging in the Si-2 filter band (8.7\,$\mu$m) and N-band (7.5--13\,$\mu$m) spectroscopy of this galaxy 
with the CanariCam on the {\it Gran Telescopio Canarias}
with a spatial resolution of 0.4--0.5$^{\prime\prime}$. 
In the {\it K'} band image the two nuclei of the galaxy each show a double-peak suggesting the complex geometry of the source, 
while the {\it L'}, {\it M'}, and Si-2 band images show
single compact structures in each of the two nuclei. 
Assuming that the center core observed at wavelengths longer than 3.8\,$\mu$m is associated with dust heated by the buried AGN, 
we re-evaluated the spectral energy distributions (SEDs) of the southern nucleus from 2 to 30\,$\mu$m with the additional literature values, and performed the SED+spectroscopy fitting using the clumpy torus models of Nenkova et al. (2008) and a Bayesian fitting approach. 
The model fit suggests that 
the high covering factor torus emission in the southern nucleus is also obscured by foreground dust in the host galaxy. 
The estimated 
AGN
bolometric luminosity of the southern nucleus, $L_{\rm bol}({\rm AGN})\sim1\times10^{45}$\,erg$\cdot$s$^{-1}$, accounts for approximately 40\% of the whole luminosity of the system.  
\end{abstract}

\section{Introduction}
Ultra-luminous and luminous infrared galaxies (ULIRGs and LIRGs) are 
galaxies characterized by their large infrared luminosities 
($L_{\rm IR} > 10^{12}L_{\odot}$ for ULIRGs and $10^{12}L_{\odot} > L_{\rm IR} > 10^{11}L_{\odot}$ for LIRGs). 
Since their discovery with the {\it IRAS} all sky survey, 
(U)LIRGs have been extensively investigated using various methods. 
They are often associated with interacting or merging galaxies (Sanders \& Mirabel 1996), 
especially at the high luminosity end. 
To date, active galactic nuclei (AGNs) buried in dust have been discovered in many (U)LIRGs 
(e.g., Imanishi et al. 2007; Veilleux et al. 2009; Nardini et al. 2010a; Imanishi et al. 2010; Alonso-Herrero et al. 2012). 
The large infrared luminosities 
of (U)LIRGs are thought to largely originate from dust heated by active starbursts and/or AGNs. 
However, various methods suggest different levels of
AGN and starburst energetic contributions in individual (U)LIRGs. 
More detailed studies of well-studied (U)LIRG are important 
to better understand the physical process happening in merger-induced (U)LIRGs.

Our target object, NGC~6240 (IRAS16504+0228), is a nearby infrared-luminous galaxy 
(z = 0.0245, Downes et al. 1993; d = 99 Mpc, 1$^{\prime\prime}$= 490 pc for $H_{0}$= 71 km s$^{-1}$ Mpc$^{-1}$, $\Omega_M$ = 0.27, $\Omega_\Lambda$ = 0.73)
with L$_{IR} = 7\times 10^{11}$L$_{\odot}$ (Wright et al. 1984; Sanders et al. 2003); 
it is classified as a LIRG, though it is on the boundary between LIRGs and ULIRGs. 
NGC~6240 is the merger of two massive and gas-rich disk galaxies 
likely in an earlier stage than typical ULIRGs (Tacconi et al. 1999). 
It has extended tidal tails and two adjacent nuclei 
separated by a projected distance of 1.5--1.7$^{\prime\prime}$ (= 740--830 pc) from north to south. 
From {\it Chandra} hard X-ray observations (Komossa et al. 2003), 
a buried AGN was discovered in each nucleus. 
Two point-like sources were also found at radio wavelengths (Gallimore \& Beswick 2004), 
providing strong additional evidence 
for
the existence of the two AGNs. 
Due to its proximity and because of its merging stage, with two nuclei hosting two buried AGNs, 
NGC~6240 is a good laboratory for investigating in detail the role of AGNs in the merger phase. 

In this paper, we present new near-infrared (NIR) and mid-infrared (MIR) observations of NGC~6240 
obtained with the {\it Subaru telescope} and the {\it Gran Telescopio Canarias} (GTC), respectively. 
We fit the data of the southern nucleus using the clumpy torus models of Nenkova et al. (2008a, b) and discuss the properties of the AGN and its possible contribution to the galaxy's total luminosity.

\section{Observations}
\begin{table*}
\caption{Log of the {\it Subaru}/IRCS+AO observations} \label{tab:1}
  \scriptsize
  \begin{center}
    \begin{tabular}{lccccccc}  \hline
Mode & Date (UT) &Filter & Plate scale & Array size & FoV & One exposure time  & Integration time  \\ \hline\hline
Imaging & 2009-08-28 & {\it K'} ($\lambda_{\rm c}$=2.12\,$\mu$m, $\Delta\lambda$=0.35\,$\mu$m) & 0.052$^{\prime\prime}$/pixel & 1024$^2$& 52$^{\prime\prime}$$\times$52$^{\prime\prime}$ & 0.41 sec & 600 sec\\
 & 2011-06-20 & {\it L'} ($\lambda_{\rm c}$=3.77\,$\mu$m, $\Delta\lambda$=0.70\,$\mu$m) & 0.020$^{\prime\prime}$/pixel & 1024$^2$& 21$^{\prime\prime}$$\times$21$^{\prime\prime}$& 0.12 sec & 130 sec\\
 & 2011-06-20 & {\it M'} ($\lambda_{\rm c}$=4.68\,$\mu$m, $\Delta\lambda$=0.24\,$\mu$m) & 0.020$^{\prime\prime}$/pixel &  768$^2$&15$^{\prime\prime}$$\times$15$^{\prime\prime}$ & 0.096 sec & 465 sec\\ \hline
    \end{tabular}
  \end{center}
\end{table*}

\begin{table*}
\caption{Log of the {\it GTC}/CanariCam observations} \label{tab:2}
  \scriptsize
  \begin{center}
    \begin{tabular}{lcccccc}  \hline
Mode & Date (UT) & Filter & Plate scale & Array size & FoV & Integration time \\ \hline\hline
 Imaging & 2013-08-27 & Si-2 ($\lambda_{\rm c}$=8.7\,$\mu$m, $\Delta\lambda$=1.1\,$\mu$m) & 0.0798$^{\prime\prime}$/pixel & 320$\times$240 & 26$^{\prime\prime}$$\times$19$^{\prime\prime}$ & 625 sec \\ \hline
Mode & Date & Grating & Slit width & Coverage & Spectral resolution & Integration time  \\ \hline
 Spectroscopy & 2013-09-15 & low-resolution 10\,$\mu$m grating & 0.52$^{\prime\prime}$ & 8--13\,$\mu$m & $R$=175 & 1238 sec \\ \hline
    \end{tabular}
  \end{center}
\end{table*}

\subsection{Subaru IRCS+AO NIR Imaging}
We carried out NIR imaging observations of NGC~6240  
with the Infrared Camera and Spectrograph 
(IRCS, Kobayashi et al. 2000) and 
the Subaru Adaptive Optics System 
(AO188, Hayano et al. 2008; Hayano et al. 2010), 
both of which are located 
at the Nasmyth focus of the
8.2\,m
{\it Subaru Telescope} on Mauna Kea in Hawaii. 
The observations were made on two different nights using three broad filters:  
{\it K'} ($\lambda_{\rm c}$=2.12\,$\mu$m, $\Delta\lambda$=0.35\,$\mu$m) on 2009 August 27 (UT), 
and {\it L'} ($\lambda_{\rm c}$=3.77\,$\mu$m, $\Delta\lambda$=0.70\,$\mu$m) 
and {\it M'} ($\lambda_{\rm c}$=4.68\,$\mu$m, $\Delta\lambda$=0.24\,$\mu$m) on 2011 June 20 (UT). 

The IRCS incorporates two 1024$^2$ ALADDIN III arrays, which are sensitive from 0.9-5.6 $\mu$m. 
The AO188 is an adaptive optics (AO) system equipped with 
a 188-element wavefront curvature sensor with photon counting avalanche photodiode modules 
and a 188-element bimorph deformable mirror.
Combined with the AO188, the IRCS achieves superb spatial resolution, 
which can be nearly equal to the diffraction limit (e.g., $\sim$0.06$^{\prime\prime}$ in the {K'} band) under ideal circumstances. 
For the present observations, 
a neighboring bright star, USNO-B1.00924-00386013 ({\it R} = 11.77 mag), 
which is located approximately 36$^{\prime\prime}$ north-east of NGC~6240, 
was used as the reference source of the AO system. 
The natural seeing of the sky was approximately 0.6--0.7$^{\prime\prime}$ in the {\it K} band. 

The plate scale and the array size were adjusted for each band filter. 
In the {\it K'} band observation, to achieve reliable sky subtraction and to investigate star clusters in the extended tidal tails, 
we opted for the 52\,mas/pixel scale and the 1024$^2$ full-array mode, 
which provides a field of view (FoV) of 52$^{\prime\prime}$$\times$52$^{\prime\prime}$. 
At wavelengths longer than $\sim$3\,$\mu$m, 
emission from the AGNs dominates the system, 
while background radiation due to the Earth's atmosphere dramatically increases. 
We therefore selected the 20\,mas/pixel scale 
with
the 1024$^2$ full-array mode 
and 
the 20\,mas/pixel scale 
with
the 768$^2$ sub-array mode 
for the {\it L'} and {\it M'} observations, respectively,
to avoid pixel saturation, focusing on a more compact area around the center of NGC~6240. 
In order to maintain a linear relationship between the incident light and the signal counts, 
we controlled the time of one exposure in each filter band. 
The times for one exposure and the total on-source integration times are listed in Table \ref{tab:1} 
together with other details of the {\it Subaru} observations.  

For the purpose of calibration, standard stars were observed on the same night at an air mass difference of 0.05 
at the beginning or end of the observation in each filter. 
We used GSPC~P565-C ({\it K'} = 11.8 mag; Hawarden et al. 2001, Leggett et al. 2006) as the standard star for the {\it K'} band, 
and SAO~140097 ({\it L'} = 6.66 mag and {\it M'} = 6.69 mag; Leggett et al. 2003) for the  standard star at the {\it Lf} and {\it Mf} bands.

The data reduction was performed in a standard manner using the IRAF package 
``ircs imgred'' (Minowa, private communication), 
which is designed for the reduction of {\it Subaru}/IRCS images.  
For the observations we used a diamond dithering pattern, 
which provided a set of nine short exposure frames with small shifts along a diamond form. 
The dithered short exposure frames were dark-subtracted, flat-fielded, and sky-subtracted with ``ircs imgred''. 
Sky flats were taken in the twilight on the same date in each filter band. 
We normalized each short exposure frame by division by the median value of the frame, 
and then median-combined them to create a master flat for each  filter band, which was applied to the whole frame. 
For the flat images, we masked out extended sources and bad pixels. 
A sky image was also generated in the same manner from each dithering set of nine frames and was applied to each frame. 
Note that we discarded some frames that were affected by severe artifacts or temporary bad weather conditions.  
We corrected for the distortion and the position angle deviation ($\sim$0.1$^{\circ}$) caused by the image rotator 
using the fits header information. 
Finally, the reduced short exposure frames were shifted and combined weighting 
by the exposure time, to create a final image. 
When we combined frames, bad pixels were masked out. 

\subsection{GTC/CanariCam MIR Imaging+Spectroscopy}
As part of the ESO/GTC AGN survey program (PI: A. Alonso-Herrero, 182.B-2005) 
with the CanariCam instrument (Telesco et al. 2003) on the GTC 10.4\,m,
we obtained an Si-2 band image centered at 8.7\,$\mu$m 
and a low-resolution N-band spectrum from 7.5 to 13\,$\mu$m of NGC~6240. 

The imaging and spectroscopic observations were performed, 
on 2013 August 27 (UT) and September 15 (UT), respectively. 
The CanariCam's 320$\times$240 Si:As detector has a plate scale of 0.0798$^{\prime\prime}$/pixel, 
which offers a FoV of 26$^{\prime\prime}$$\times$19$^{\prime\prime}$ in the imaging mode.

We obtained an image of NGC~6240 using the Si-2 filter, which has
a central wavelength of $\lambda_{\rm c}$=8.7\,$\mu$m 
and a width of $\Delta\lambda$=1.1\,$\mu$m at 50\% cut-on/off. 
We used chopping and nodding throws of 15$^{\prime\prime}$, 
with chop and nod position angles of -74$^{\circ}$ and 106$^{\circ}$, respectively. 
These were chosen to avoid extended emission of the galaxy, 
while ensuring good image quality.  
Immediately after the galaxy observation we obtained an image of the standard star HD151217 in the same filter, 
which was used to perform the photometric calibration. 
The FWHM of the resulting images was 0.38$^{\prime\prime}$ in the standard star observation.

For the spectroscopic observation, 
we used the low spectral resolution 10\,$\mu$m grating, 
which covers the N-band (7.5--13\,$\mu$m) 
with a nominal resolving power of $R$=175. 
A 0.52$^{\prime\prime}$ wide slit was oriented at a position angle of 16$^{\circ}$ east of north so that both nuclei were imaged on the slit. 
The observing sequence was as follows. 
We first obtained an acquisition image of NGC~6240 using the Si-2 filter, 
then we placed the slit and integrated for a total on-source time of 1238 seconds. 
The chop and nod parameters were the same as for the imaging observations.
The standard star HD15799
was observed just after the target observation 
following the same observing sequence. 
We performed the photometric calibration and the telluric correction, and computed the slit losses 
using the standard star observation. 
The image quality of the spectroscopic observations, measured from the standard star, was 0.40$^{\prime\prime}$ (FWHM). 

The observational parameter, 
for the imaging and spectroscopy of NGC~6240 with the GTC/CanariCam,
including the total on-source integration time, 
are summarized in Table \ref{tab:2}. 
The data reduction was done 
using the CanariCam pipeline RedCan (Gonz\'{a}lez-Mart\'{i}n et al. 2013 for a full description). 
For the spectroscopic data we extracted the spectrum of the southern nucleus as a point source
as the emission from this nucleus appears unresolved at the angular resolution of the observations. 
The northern nucleus appears extended in the MIR (see section 3.2) and therefore we 
do
not attempt to fit it with the clumpy torus models. 
For more details in the GTC/CanariCam data reduction and analysis, see Alonso-Herrero et al. (2014).

\begin{figure*}
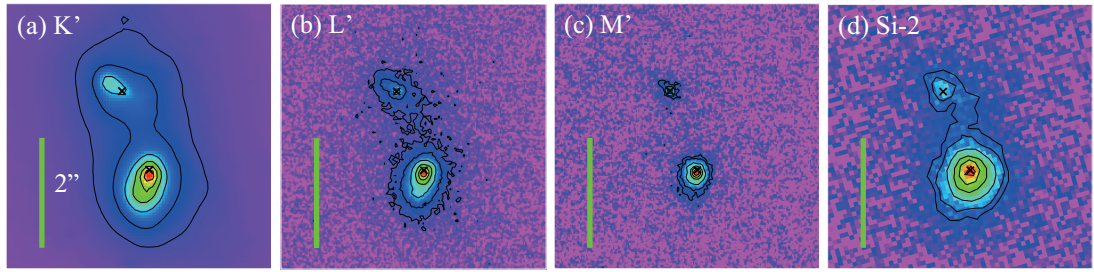

  \begin{center}
\FigureFile(411.3pt,163.3pt){fig1_revised.eps}
  \end{center}
  \caption{The {\it Subaru}/IRCS+AO {\it K'}, {\it L'}, and {\it M'} band images (a)$\sim$(c), 
  and the {\it GTC}/CanariCam Si-2 band image (d). 
  North is up and east is left. All images are displayed at the original angular resolution. 
  Crosses indicate the position of the two nuclei on the radio map at 1.7\,GHz (Gallimore \& Beswick 2004), 
  if we assume that the position of the northern nucleus in each image coincides with that of radio. 
  The value of the lowest contour is 200$\sigma$ in the {\it K'} band image, 
and 3$\sigma$ in the {\it L'}, {\it M'}, and Si-2 band images.}
  \label{fig:1}
\end{figure*}

\begin{table*}
\begin{minipage}{\hsize}
\caption{Aperture photometry of the northern and southern nuclei \footnotemark[1]}\label{tab:3}
  \scriptsize
  \begin{center}
    \begin{tabular}{lcccc}  \hline
Name&Aperture Diameter& K'&  L'&  M' \\ \hline\hline
North 2  & \,\,\,\,0.34$^{\prime\prime}$ (= 0.17 kpc) \footnotemark[2] & \,\,\,0.80$\pm$0.01 (14.8)    &  \,\,\,1.06$\pm$0.21 (13.4)    &   \,\,\,2.13$\pm$0.45 (12.2) \\
              & \,\,\,1.0$^{\prime\prime}$ (= 0.49 kpc)                               & \,\,\,5.41$\pm$0.05 (12.7)    &  \,\,\,3.67$\pm$0.11 (12.1)  &   \,\,\,4.37$\pm$0.23 (11.4) \\
              & \,\,\,1.1$^{\prime\prime}$ (= 0.54 kpc)                                & \,\,\,6.06$\pm$0.06 (12.6) & &     \\
              & \,\,\,5.0$^{\prime\prime}$ (= 2.45 kpc)                               & 46.78$\pm$0.47 (10.4) & &     \\
              & 11.4$^{\prime\prime}$ (= 5.59 kpc)                                    & 68.41$\pm$0.68 (10.0) & &     \\ \hline
South 1  & \,\,\,\,0.34$^{\prime\prime}$ (= 0.17 kpc) \footnotemark[2] & \,\,\,6.22$\pm$0.08 (12.6)   &    \,\,\,6.65$\pm$1.37 (11.4)    &  15.82$\pm$3.20 (10.0) \\
              & \,\,\,1.0$^{\prime\prime}$ (= 0.49 kpc)                             & 16.02$\pm$0.16 (11.6)   &  15.44$\pm$0.34 (10.5) &   21.86$\pm$0.58 (9.68) \\
              & \,\,\,1.1$^{\prime\prime}$ (= 0.54 kpc)                               & 17.34$\pm$0.17 (11.5)   & &     \\
              & \,\,\,5.0$^{\prime\prime}$ (= 2.45 kpc)                               & 47.11$\pm$0.47 (10.4)   & &     \\
              & 11.4$^{\prime\prime}$ (= 5.59 kpc)                                   & 67.95$\pm$0.68 (10.0)  & &     \\ \hline
Center   & \,\,\,4.0$^{\prime\prime}$ (= 1.96 kpc)                                & 43.39$\pm$0.43 (10.5)   &  30.99$\pm$0.77 (9.76)   &  33.81$\pm$0.12 (9.21)\\ \hline
    \end{tabular}
  \end{center}
\footnotetext[1]{Measured in mJy. Values in a unit of Vega magnitude are also described in parentheses. See the detailed information about the flux densities for Vega for the Mauna Kea Observatories Near-Infrared filter set and the conversion of the units in Tokunaga \& Vacca (2005).}
\footnotetext[2]{Aperture-corrected and stellar-subtracted.}
\end{minipage}
\end{table*}

\begin{table*}
\begin{minipage}{\hsize}
\caption{MIR photometric data of the South 1\footnotemark[1]}\label{tab:4}
  \scriptsize
  \begin{center}
    \begin{tabular}{lcccc}  \hline
Name & Wavelength ($\mu$m)& Egami et al. (2006) &  estimated starburst flux &   starburst-subtracted flux \\ \hline\hline
South 1 & 7.9&  275.0$\pm$   27.5&   50.0$\pm$   10.3&  225.0$\pm$   29.4\\
& 8.8&  142.0$\pm$   14.2&   26.9$\pm$    5.5&  115.1$\pm$   15.2\\
&10.3&   67.0$\pm$    6.7&    7.9$\pm$    1.8&   59.1$\pm$    6.9\\
&11.7&  201.0$\pm$   20.1&   32.6$\pm$    7.0&  168.4$\pm$   21.3\\
&12.5&  375.0$\pm$   37.5&   42.1$\pm$    9.0&  332.9$\pm$   38.6\\
&17.9&  919.0$\pm$   91.9&   83.2$\pm$   17.3&  835.8$\pm$   93.5\\
&24.5& 2734.0$\pm$  273.4&  206.1$\pm$   42.2& 2527.9$\pm$  276.6\\ \hline
    \end{tabular}
  \end{center}
\footnotetext[1]{Measured in mJy.}
\end{minipage}
\end{table*}

\section{Results}
\subsection{Subaru NIR Data}
The {\it Subaru}/IRCS+AO {\it K'}, {\it L'}, and {\it M'} band images are presented in Figures 1a, b, and c, respectively. 
In the {\it K'} band image, 
the FWHM of a point-like source detected in the vicinity of the center of NGC~6240 is approximately 0.17$^{\prime\prime}$. 
Although the AO correction varies 
depending on the distance from the reference star,  
it can be considered that at least in the area that we focus on the spatial resolution is comparable to 0.17$^{\prime\prime}$. 
On the other hand, in the {\it L'} and {\it M'} band images, we do not have an accurate probe,  
since no source was detected with the exception 
of
NGC~6240. 
However, from the standard observations, 
we can estimate that the image quality is at most 0.10$^{\prime\prime}$ (FWHM) in the {\it L'} band and 0.12$^{\prime\prime}$ (FWHM) in the {\it M'} band. 
In the {\it M'} band image, the two nuclei are observed as essentially point sources 
(FWHM=0.12$^{\prime\prime}$ for the northern nucleus and FWHM=0.15$^{\prime\prime}$ for the southern nucleus). 
In the {\it L'} band image, although there are extended components arising from the host galaxy, 
we can also see the core structure in each nucleus, similar to that of the {\it M'} band image.

As reported in Max et al. (2005) and Pollack et al. (2007), 
in which the images had resolution more like 0.05$^{\prime\prime}$, 
the two nuclei are elongated, each exhibiting a double-peak in the {\it K'} band image. 
Although it also shows compact core structures, 
much of the {\it K'} emission in the nuclear region is due to photospheric emission. 
In contrast, in addition to being compact, 
the angular separation of the {\it L'} and {\it M'} nuclear sources is nearly the same as that of the radio peaks, which mark the location of the active nuclei. 
The observed trend can be interpreted as emission from AGNs in the two nuclei 
being dominant against stellar emission for increasing wavelength. 
Pollack et al. (2007) and Max et al. (2007) suggest that 
the sub-peak of the northern nucleus in the {\it K'} band image, ``North 2'', is coincident with the northern black hole.
In Figure \ref{fig:1}, we plot the position of the two nuclei on the radio map at 1.7\,GHz (Gallimore \& Beswick 2004) with cross points,  
assuming that 
the North 2 source in the {\it K'} band and the northern peak observed at wavelengths longer than 3.8\,$\mu$m is spatially coincident with the radio source. 
As shown by the figures, the angular separation and the position angle between the two nuclei in the {L'} and {M'} band 
closely match those detected in the radio. 
This result confirms that the two compact structures detected in the {\it L'} and {\it M'} images are associated with emission from the AGNs located in the northern and southern nuclei. 
Hereafter, the northern and southern nucleus are referred to  as ``North 2'' and  ``South 1'' respectively, 
following Max et al. (2005), Pollack et al. (2007), and Max et al. (2007).

The PSF of an AO 
system is composed of a core and a halo component. 
We therefore performed aperture photometry on the North 2 and South 1
to estimate the flux densities associated with the buried AGN in each nucleus. 
The aperture is a circle with a diameter of 0.34$^{\prime\prime}$ (= 0.17 kpc), 
and the background is an annulus with a mean diameter of 0.425$^{\prime\prime}$ (= 0.40 kpc) and a width of 0.085$^{\prime\prime}$ (= 0.04 kpc). 
The total flux was calculated from the measured counts using a conversion factor, 
which is the ratio of the counts measured with a particular aperture and background annulus to the calibrated total flux. 
In the {\it K'} band, the conversion factor was estimated from a point-like source, 
which was detected in the region surrounding the center of NGC~6240 
at the coordinates J2000 16:52:58.68, +02:23:56.9 (RA, Dec). 
Its total flux was calculated from the counts measured with a larger aperture using an ADU-to-flux ratio, 
which was determined through calibration with a standard star. 
In the case of the {\it L'} and {\it M'} band images, 
for lack of usable sources, 
the conversion factor was estimated from the standard star observations. 
In the {\it L'} and {\it M'} bands, the standard star was observed with the assistance of AO 
using the same control regime as for the object, 
except for the brightness of the guide star and its separation from the target object. 
We estimated the uncertainty due to the differences to be 20\% on the basis of the analysis of Imanishi \& Saito (2014), 
and included it in the errors.  
Table \ref{tab:3} provides a summary of the aperture photometry
in units of mJy and Vega magnitude. 
The fluxes measured with larger apertures are also tabulated for reference. 
The {\it K'} band fluxes measured with 1.1$^{\prime\prime}$ (= 0.54 kpc), 5.0$^{\prime\prime}$ (= 2.45 kpc), and 11.4$^{\prime\prime}$ (= 5.59 kpc) aperture diameters 
are in good agreement with the photometric results of Scoville et al. (2000), 
and the discrepancy between two sets of measurements are within 10\% of the absolute fluxes. 

\subsection{GTC MIR Data}
The {\it GTC}/CanariCam Si-2 band image is shown in Figure 1d. 
As mentioned in section 2.2, the image quality 
as measured from the standard star is approximately 0.38$^{\prime\prime}$ (FWHM). 
Egami et al. (2006) obtained the 7.9, 8.8, 10.3, 11.7, 12.5, 17.9, and 24.5\,$\mu$m images of NGC~6240 
with the MIRLIN camera (Ressler et al. 1994) on the {\it Keck} telescope 
and measured the flux densities of the two nuclei within a circle of a 1$^{\prime\prime}$ (= 0.49 kpc) diameter at each wavelength. 
As 
was
also reported in Egami et al. (2006), 
the North 2 seems extended, 
whereas the South 1 is unresolved, 
although the peak position of the two nuclei are in good agreement with 
that of the {\it L'} and {\it M'} band images obtained with {\it Subaru}/IRCS+AO. 
The extended nature of the MIR emission of the North 2 indicates that contribution from dust heated by star formation activity is higher than or comparable with that of dust in the AGN torus.
This is also confirmed by the strong 11.3\,$\mu$m PAH emission arising from this nucleus 
(see Alonso-Herrero et al. 2014 for a detailed analysis of the PAH emission of this galaxy). 

We performed aperture photometry on the GTC/CanariCam Si-2 band image with a 1$^{\prime\prime}$ diameter 
and extracted a spectrum as a point source for the South 1. 
After correction for the point source, the 1$^{\prime\prime}$ diameter aperture photometry gave a flux of 163\,mJy at 8.7\,$\mu$m. 
This value is consistent with the flux density at 8.8\,$\mu$m reported by Egami et al. (2006). 

The GTC/CanariCam spectrum of the South 1 
shows PAH emission at 8.6 and $11.3\,\mu$m 
(see Alonso-Herrero et al. 2014 for a detailed discussion) 
and a deep silicate feature coming from an unresolved region of approximately 240\,pc in diameter. 
We measured the apparent strength of the $9.7\,\mu$m silicate feature 
as $S_{\rm Si}=\ln (f_{\rm obs}(9.7\mu{\rm m})/f_{\rm con}(9.7\mu{\rm m})$), 
where $f_{\rm obs}(9.7\mu{\rm m})$ is the observed flux density of the feature and 
$f_{\rm con}(9.7\mu{\rm m})$ is the continuum flux density. 
The limited spectral range of the GTC/CanariCam spectroscopy makes it difficult 
to estimate the continuum above the silicate feature. 
We used the {\it Spitzer}/IRS spectrum to obtain the slope of the continuum 
by fitting it linearly between 5 and $15\,\mu$m. 
We then assumed that 
the slope of the fitted continuum in the {\it Spitzer}/IRS is the same as that of the GTC/CanariCam spectrum. 
We obtained a nuclear strength of the silicate feature $S_{\rm Si}=-1.50$, 
which is typical of deeply embedded ULIRGs (e.g., Spoon et al. 2007). 

\subsection{Model fitting}
In order to quantitatively estimate the AGN properties, 
we fitted the spectral energy distribution (SED) plus 
nuclear spectroscopy 
of the South 1 using the clumpy torus models of  Nenkova et al. (2008a, b). 
These models were found to reproduce the nuclear infrared emission of Seyfert  galaxies and PQ quasars 
(Ramos Almeida et al. 2009, 2011; Mor et al. 2009; Alonso-Herrero et al. 2011; Lira et al. 2013). 
As discussed in the previous section, 
the MIR emission of the North 2 
appears to be dominated by star formation activity. 
We therefore did not attempt to fit it with the clumpy torus models. 

In the fit, in addition to the present {\it Subaru}/IRCS+AO NIR photometry data 
and the {\it GTC}/CanariCam MIR spectroscopic data, 
we utilized the published data from Egami et al. (2006) and Armus et al. (2006) as a supplement. 
Egami et al. (2006) suggested that the MIR spectrum of NGC~6240 is mostly dominated by emission from dust heated by starburst activity and does not show an apparent signature of the hidden AGN. 
To account for this, we re-evaluated the contribution from starburst emission at these MIR wavelengths 
using the {\it Subaru} {\it L'} band photometric results 
and the {\it ISO}/SWS spectrum of the well-studied starburst galaxy M82 (Sloan et al. 2003). 
We first considered the difference between the 1$^{\prime\prime}$ diameter flux and the flux associated with the buried AGN, 
which was estimated from aperture photometry (see details in Section 3.1) to be the starburst component in the {\it L'} band. 
Subsequently, the flux densities associated with the starburst component at MIR wavelengths were extrapolated 
based on the assumption that the starburst component has the same spectral distribution as M82 from the NIR to MIR, 
and these were subtracted from the data. 
We added statistical errors to a typical 10\% photometric uncertainty in quadrature to compute the error. 

The original values of Egami et al. (2006) and the corrected values are summarized in Table \ref{tab:4}, 
together with the estimated starburst fluxes. 
As suggested by Brandl et al. (2006), 
compared with the differences among 
different kinds of astronomical objects, 
the variance of the IR SEDs among starburst galaxies is relatively small.
Using the Spitzer/IRS spectra of starburst samples provided by Hern\'an-Caballero \& Hatziminaoglou (2011), 
we estimate the variation of SB SEDs from NIR to MIR 
as within a factor of ~2 of the absolute value of estimated starburst fluxes. This suggests that
even after the subtraction of the starburst component, 
there remains significant emission, 
which 
can not
be explained only by the starburst SED scatter, and 
we assume that it comes from dust in the torus heated by the AGN. 

In Armus et al. (2006), the MIR spectrum of NGC~6240 taken with the Infrared Spectrograph (IRS) on the {\it Spitzer Space Telescope} were presented. 
We took its monochromatic flux at 30\,$\mu$m as an upper limit to the AGN emission, 
since the slit width of the IRS long wavelength module is 
approximately 10$^{\prime\prime}$ and it covered both nuclei.  
For consistency, the {\it GTC}/CanariCam MIR spectrum was scaled to match 
the 8.8\,$\mu$m starburst-subtracted flux of Egami et al. (2006), and 
was rebinned to 40-50 data points, 
which was found to be the optimal number in the spectra for the fitting 
(see Alonso-Herrero et al. 2011 for details). 
In order to avoid spectral regions with low 
signal-to-noise
ratios 
being
included in the model fit, 
we trimmed the edges of the spectrum.
For the error estimation, 
we assumed a 15\% photometric and/or scaling uncertainty 
plus the error computed from the rms of the spectrum. 
Judging from the SED shape, 
although it is after local background subtraction, 
the {\it Subaru} {\it K'} band data point 
still seems 
definitely contaminated with stellar emission. 
This implication is also supported by the fact that 
even if we include it as a detection in the model fitting, 
the fitted models are well below the {\it K'} band data point, 
and we therefore took it as an upper limit in the following.

\begin{table*}
\begin{minipage}{\hsize}
\caption{Ranges of the Nenkova et al. (2008a, b) clumpy torus model parameters}\label{tab:5}
  \scriptsize
  \begin{center}
    \begin{tabular}{lcccccc}  \hline
Parameter & {\it $\sigma_{\rm torus}$} (deg) &  {\it Y} & {\it $N_0$}  & {\it q} & {\it $\tau_{\rm V}$} &  {\it i} (deg)  \\ \hline\hline
Interval &[15, 70] & [5, 100] & [1, 15] &  [0, 3] &  [5, 150] & [0, 90] \\ \hline
    \end{tabular}
  \end{center}
\end{minipage}
\end{table*}

{
\scriptsize
\begin{table*}
\begin{minipage}{\hsize}
\begin{center}
\caption{Fitted clumpy torus model parameters}\label{tab:6}
\vskip 0.1pt
\scalebox{0.65}[0.65]
{
   \begin{tabular}{cccccccccccccccccccc}
\hline
\multicolumn{1}{c}{Parameter}&
\multicolumn{2}{c}{{\it $\sigma_{\rm torus}$}(deg)}&&
\multicolumn{2}{c}{\it Y}&&
\multicolumn{2}{c}{\it $N_0$}&&
\multicolumn{2}{c}{\it q}&&
\multicolumn{2}{c}{\it $\tau_{\rm V}$}&&
\multicolumn{2}{c}{{\it i} (deg)}\\
\cline{2-3}\cline{5-6}\cline{8-9}\cline{11-12}\cline{14-15}\cline{17-18}
&
\multicolumn{1}{c}{Median}&\multicolumn{1}{c}{MAP}&&
\multicolumn{1}{c}{Median}&\multicolumn{1}{c}{MAP}&&
\multicolumn{1}{c}{Median}&\multicolumn{1}{c}{MAP}&&
\multicolumn{1}{c}{Median}&\multicolumn{1}{c}{MAP}&&
\multicolumn{1}{c}{Median}&\multicolumn{1}{c}{MAP}&&
\multicolumn{1}{c}{Median}&\multicolumn{1}{c}{MAP}\\
\hline\hline
South 1&
 64$_{-5}^{+4}$& 70 &&
 29$_{-4}^{+7}$&24&&
 12$^{-2}_{+2}$&15&&
0.39$_{-0.24}^{+0.33}$& 0.06 &&
 41$_{-10}^{+13}$ & 49  &&
 50$_{-16}^{+18}$ & 26 \\ 
\hline
\end{tabular}
}
\end{center}
\end{minipage}
\end{table*}
}

In this study, we used the new version of BayesClumpy and the interpolated version of the models of Nenkova et al. (2008a, b) 
as explained in Asensio Ramos \& Ramos Almeida (2009)
and Asensio Ramos \& Ramos Almeida (2013). 
The new version of BayesClumpy now uses a linear interpolation of the torus models. 
The model parameters are as follows: 
the torus angular width {\it $\sigma_{torus}$}, 
the torus radial thickness {\it Y} that is defined as the ratio between the outer and inner radii of the torus, 
the mean number of clouds {\it $N_0$} along an equatorial ray, 
the power-law index {\it q} of the radial density distribution of the clouds, 
the optical depth {\it $\tau_V$} per single cloud, 
and the viewing angle {\it i}. 
We did not fix any of them, and their prior distributions were assumed to be uniform within the model range. 
Their prior ranges are presented in Table \ref{tab:5}. 
Within the BayesClumpy routine there are three extra parameters we can use for the fit. 
The first one is the vertical shift to match a torus model to an SED. 
This is the scaling of a model spectrum, 
which is directly related to the bolometric flux of an AGN. 
We took it as a free parameter so that 
we could estimate the AGN bolometric luminosity of the South 1 from the fit. 
The second one is the foreground extinction due to the host galaxy. 
We assumed foreground extinction with a Gaussian distribution centered at $A_{\rm V}=7$\,mag with a width of 4\,mag as a prior, based on Pollack et al. (2007). 
The final parameter was the redshift, which was fixed to 0.0245. 

\begin{figure}[htbp]
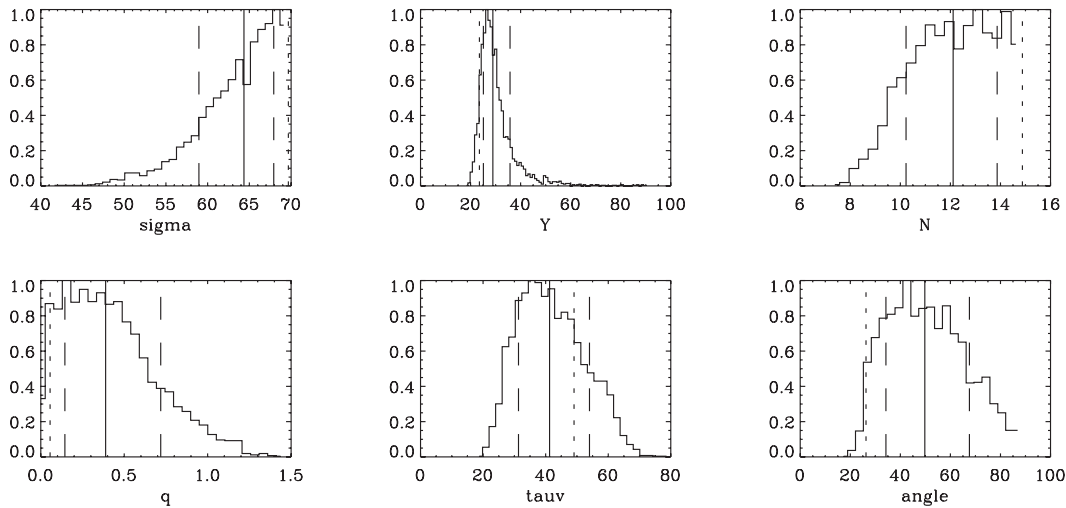

  \begin{center}
    \FigureFile(400pt,200pt){fig2_revised.eps}
  \end{center}
    \caption{The marginal posterior distributions of the fitted model parameters. 
The median and MAP values of each of the Nenkova et al. (2008a,b) torus model parameters
are indicated by solid and dashed lines respectively. 
    The long-dashed lines illustrate one-sigma uncertainty of the fit. }
    \label{fig:2}
\end{figure}

\begin{figure}
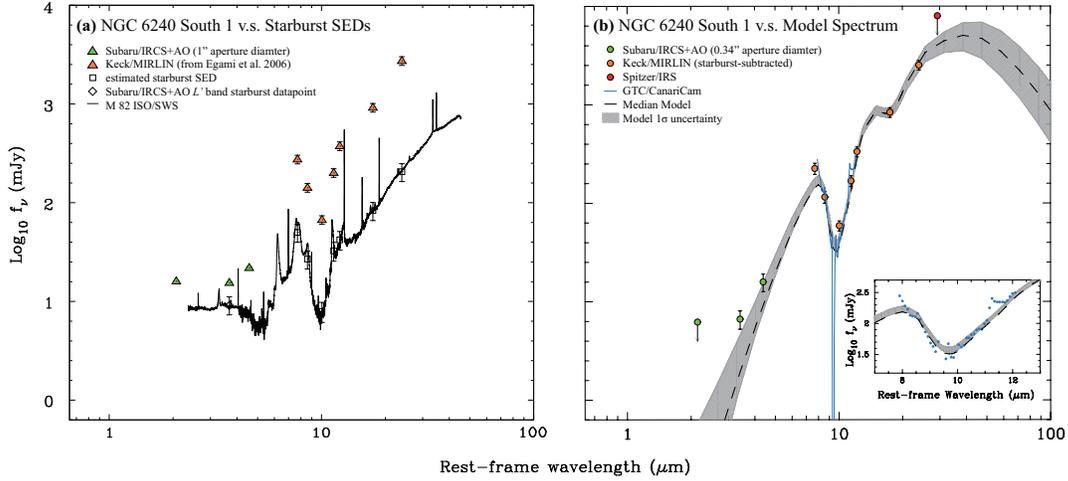

  \begin{center}
    \FigureFile(400pt, 200pt){fig3_revised.eps}
   \end{center}
    \caption{
{\bf(a)} The 1$^{\prime\prime}$ aperture diameter fluxes of the South 1 (triangles)
v.s. the estimated starburst fluxes (squares). 
The {\it Subaru}/IRCS+AO {\it K'}, {\it L'}, and {\it M'} band fluxes
and {\it Keck}/MIRLIN photometric data points from Egami et al. (2006)
are represented by green and orange, respectively. 
The {\it ISO}/SWS spectrum of the starburst galaxy M~82, 
which was used in the estimation of the starburst contribution at the MIR wavelengths (see details in section 3.3), 
is scaled and shown with the black line. 
{\bf(b)} The fitted SEDs and spectrum of the South 1
v.s. the best-fit Nenkova et al. (2008a,b) clumpy torus model. 
The filled circles denote the photometric data points: 
the {\it Subaru}/IRCS+AO {\it K'}, {\it L'}, and {\it M'} band 0.34$^{\prime\prime}$ aperture diameter fluxes (green), 
the starburst-subtracted {\it Keck}/MIRLIN fluxes (orange), 
and the {\it Spitzer}/IRS monochromatic flux at 30\,$\mu$m from Armus et al. (2006, red). 
The blue line indicates the GTC/CanariCam nuclear spectrum. 
The downward arrows attached to the {\it Subaru} {\it K'} band and the {\it Spitzer} 30\,$\mu$m data points 
indicate that they are taken as upper limits in the model fit. 
The dashed line and the grey shaded area represent the median fitted model spectrum and one-sigma uncertainty of the fit. 
The inset shows the GTC/CanariCam nuclear spectrum (blue dots) and the fitted torus model (line and shaded region). 
The excess emission with respect to the torus model fit is the nuclear 11.3\,$\mu$m PAH emission.
}
    \label{fig:3}
\end{figure}

Figures \ref{fig:2} and \ref{fig:3} show the fitting results, 
the marginal posterior distributions for the clumpy torus model parameters and 
the model spectrum together with the observed SED and spectrum. 
As shown by Figure \ref{fig:3}, 
the best fit model reproduces the observed SEDs and the spectrum fairly well. 
The excess emission in the GTC/CanariCam spectrum not fitted by the torus models 
(see inset of Figure \ref{fig:3}) 
is due to the nuclear 11.3\,$\mu$m PAH feature in the South 1 
(see Alonso-Herrero et al. 2014 for a detailed discussion). 
It is important to note that the fit requires 
a relatively high value of foreground extinction $A_{\rm V}$= 19$\pm$3 mag. 
The high value of the foreground extinction 
needed to fit the deep nuclear silicate feature ($S_{\rm Si}=-1.50$, see Section 3.2) 
of the South 1 of NGC 6240 
is because clumpy torus models (including that of Nenkova et al. 2008a,b) 
never produce deep silicate features 
(see Levenson et al. 2009 and Gonz\'alez-Mart\'{\i}n et al. 2013). 
In particular, the database of the Nenkova et al. (2008a,b) models shows 
a $3\sigma$ range of all $S_{\rm Si}$ values (99.7\%) [-1.45, +1.19], 
with a median value at 0.30. 
The fitted value of the foreground extinction is higher than 
the average value of $A_V=7\,$mag 
inferred by Pollack et al. (2007) for the star clusters in NGC~6240, 
although for individual clusters near the nuclei 
they derived values of the extinction as high as $A_V=14\,$mag. 
On the other hand, from the modeling of the stellar population 
of the nuclear spectral energy distribution of this nucleus, 
Engel et al. (2010) estimated an effective extinction 
in the $K$-band of approximately  $A_K=8\,$mag at the location of the South 1, 
which would be equivalent of an optical extinction of $A_V\sim 80\,$mag.

The median and maximum a posteriori (MAP) values and 
their associated one-sigma uncertainties of the model parameters 
are given in Table \ref{tab:6}. 
As can be seen from Table \ref{tab:6}, the data require 
a high value of the torus angular size $\sigma_{\rm torus}$ and number of clouds along the equatorial direction {\it $N_0$}, 
which results in a torus with a high covering factor 
$f_2=0.97^{+0.03}_{-0.02}$
(see also Figure 10 of Ramos Almeida et al. 2011). 
This together with the required high value of the foreground extinction 
are similar to the findings for the fitted torus for NGC~3690, 
which is one of the nuclei of the interacting LIRG Arp~299 
(Alonso-Herrero et al. 2013). 
Using photometric and spectroscopic IR data of Seyfert galaxies, 
Ramos Almeida et al. (2014) showed that 
for both 
Seyfert
1 and 
Seyfert
2 galaxies 
{\it J}+{\it K}+{\it M} band photometry and {\it N} band spectroscopy is needed 
to reliably constrain the torus parameters in general, 
and, when focusing on Seyfert 2 galaxies in particular, 
the minimum required filter set is {\it H}+{\it M} band photometry and {\it N} band spectroscopy. 
In this case, at $\lambda <3\,\mu$m, 
we only have an upper limit to the unresolved flux at 2.2\,$\mu$m. 
However, the Subaru/IRCS+AO {\it L'} and {\it M'} band photometric data points seem to be enough for determining the NIR-to-MIR slope, 
and the GTC/CanariCam {\it N} band spectrum 
reduces
the parameter space significantly. 
We note that even if 
we treat the {\it K'} band flux as a detection instead of an upper limit, 
the fitted models show almost no changes and accordingly the obtained value of the covering factor $f_2$ is very similar to 
that obtained using the 
{\it K'} band
flux as an upper limit.

\section{Discussion and Conclusion}

The AGN bolometric luminosity and its contribution in NGC~6240 have been measured in various ways to date. 
Komossa et al. (2003) reported that the 0.1-10\,keV absorption-corrected luminosities of the north and south nuclei are 
$3\times 10^{41}$ and $1\times 10^{42}$\,erg$\cdot$s$^{-1}$ respectively (corrected to our distance). 
However, as they point out, their estimation is only a lower limit on the intrinsic luminosity of the AGN, 
since the flux below 10\,keV originates from scattered emission. 
Lutz et al. (2003) constrained the AGN and starburst contributions in NGC~6240 via multiple methods 
using {\it ISO}/SWS+LWS spectroscopy, 
and concluded that the AGN contribution is in the range of 25--50\%.  
Armus et al. (2006) estimated the AGN contribution at around 20--24\% with the {\it Spitzer}/IRS spectroscopy. 
Veilluex et al. (2009) also comprehensively investigated the contribution of the nuclear activity to the whole system 
for 34 PG quasars and 74 ULIRGs with z = $\sim$0.3 including NGC~6240  
based on the {\it Spitzer}/IRS spectroscopy using six independent methods. 
In their analysis, the average derived AGN contribution in NGC~6240 was 25.8\%. 
From the view of ground-based observations, 
Egami et al. (2006) suggested that 
the AGN bolometric luminosity can be at most 60\% of the whole energy of the galaxy, 
if it is heavily obscured.

In this study, 
combining the MIR data with the NIR high-spatial-resolved imaging data, 
we re-evaluated the SEDs and the spectrum of 
the South 1 of this galaxy 
and fitted these 
data
with the Nenkova et al. (2008a,b) torus models. 
This is an improvement over previous studies of this galaxy in the infrared 
where the AGN emission was not fitted with torus models. 
From the present model fitting, we inferred that 
the AGN in the South 1 has a high covering factor torus 
and it is deeply embedded in the host galaxy with $A_{\rm V}$$\sim$19 mag. 
NGC~6240 is classified as a LINER rather than a Seyfert in the optical (Armus et al. 1989), 
whereas the [Ne V] line is detected in the infrared (Armus et al. 2006). 
This is likely indirect evidence of high extinction toward the nuclear regions of NGC~6240.  
Besides, the X-ray observations indicated large hydrogen column densities toward the nuclei (Komossa et al. 2003). 
These results also support a high covering factor torus suggested by the model fit.
The estimated bolometric luminosity of the AGN is $1.3\times10^{45}$\,erg$\cdot$s$^{-1}$. 
This corresponds to the median value of the scaling parameter. 
The one-sigma uncertainty of the derived probability distribution of the scaling parameter is 0.10 dex. 
Yet for Seyfert nuclei, 
the typical uncertainty in estimating the AGN bolometric luminosity from the torus fitting is estimated at approximately 0.4 dex 
when 
compared 
with other estimates in the literature (see Alonso-Herrero et al. 2011, for more details). 
Alternatively, we can estimate the AGN bolometric luminosity 
using the local mid-IR vs. X-ray correlation in AGN (e.g., Asmus et al. 2011). 
Using the starburst-subtracted $12.5\,\mu$m flux density measurement 
of the South 1 (Table \ref{tab:4}) 
we estimated an intrinsic $2-10$\,keV luminosity of $5\times 10^{43}\,{\rm erg\,s}^{-1}$. 
After applying a bolometric correction to the hard X-ray luminosity (Marconi et al. 2004) 
we would obtain an AGN bolometric luminosity 
for the South 1 similar to that obtained from the torus model fitting.

The derived bolometric luminosity of the southern AGN 
contributes 37\% of the total energy of the galaxy, 
if we assume that the infrared luminosity is almost the same as the total luminosity of the system. 
Here we do not perform any model fit for the North 2. 
However, considering that the flux ratios between the two nuclei are around 
7--8 at IR wavelengths and 2--3 at radio and hard-Xray wavelengths, 
it can be inferred that the AGN contribution from the two nuclei is in the range of 45--60\% in total, 
which is in good agreement with the previous estimations. 
Alonso-Herrero et al. (2012) showed that 
in local LIRGs the AGN bolometric contribution to the IR luminosity of the system is generally small ($<\sim$5\%). 
On the other hand, it is known that in more luminous LIRGs and especially in ULIRGs 
the AGN contribution tends to be much higher (Nardini et al. 2008, 2010).
These results corroborate that in the local universe 
the AGN contribution increases with the IR luminosity of the system 
(Nardini et al. 2010; Imanishi et al. 2010a,b; Alonso Herrero et al. 2012). 
The high value of the derived AGN contribution is consistent with 
the fact that NGC~6240 sits on the boundary between LIRGs and ULIRGs 
with L$_{\rm IR} = 7\times 10^{11}$L$_{\odot}$ (Wright et al. 1984; Sanders et al. 2003).

Despite the relatively large uncertainty in the estimated AGN bolometric luminosities, 
the present result suggests that 
in the phase of mergers 
even though the observed AGN luminosities are low, 
the intrinsic energy contribution of the AGNs to the whole system is not a negligible amount. 
This is because both AGN  are buried deeply in the host galaxy. 
It may possibly indicate that 
the growth of 
the
dust torus is associated with the galaxy evolution due to the galaxy merger. 
In the {L'} and {M'} bands, 
the contribution of emission from stars and dust such as PAHs is relatively small. 
As shown in this study, 
the {L'} and {M'} band high-spatial-resolved observations with the assist of AO 
in combination with diffraction-limited MIR observations 
will be useful for the investigation of such a hidden dusty AGN in other galaxies.

\section*{Acknowledgements}
We would like to thank the anonymous referee for his or her useful comments, 
which improved the paper significantly.
We would like to express our gratitude to Dr. Minowa 
for his continuing support during and after our observations at the {\it Subaru Telescope}, 
and Aaron C. Bell for his English proofreading. 
T.~I.~M. receives financial support from a Grant-in-Aid for Japan Society for the Promotion of Science (JSPS) Fellows. 
M.~I. is supported by Grants-in-Aid for Scientific Research (no. 23540273). 
A.~A.-H. acknowledges support from the Universidad de Cantabria through the Augusto G. Linares programme and
from the Spanish Plan Nacional grants AYA2009-05705-E and AYA2012-31447. 
C.~P. acknowledges support from NSF-0904421. 
C.~R.~A. is supported by a Marie Curie Intra European Fellowship within the 7th European Community Framework Programme (PIEF-GA-2012-327934). 
R.~N. acknowledges support by the ALMA-CONICYT fund, project No. 31110001, and by FONDECYT grant No. 3140436.
N.~A.~L. is supported by the Gemini Observatory, 
which is operated by the Association of Universities for Research in Astronomy, Inc., 
on behalf of the international Gemini partnership of Argentina, Australia, Brazil, 
Canada, Chile, and the United States of America.
This work is based on observations made with the GTC, 
installed in the Spanish Observatorio del Roque de los Muchachos of the Instituto de Astrof\'{i}sica de Canarias, 
in the island of La Palma, and the {\it Subaru Telescope}. 
Part of the {\it Subaru} data were taken in the framework of 2009 Subaru observing event for undergraduate students in Japan. 
We would like to offer our special thanks to all the persons who were involved in this event.


\end{document}